\documentclass[conference]{IEEEtran}
\IEEEoverridecommandlockouts
\usepackage{cite}
\usepackage{amsmath,amssymb,amsfonts}
\usepackage{algorithmic}
\usepackage{graphicx}
\usepackage{textcomp}
\usepackage{xcolor}
\usepackage{algorithm}
\usepackage{color, soul, colortbl}
\usepackage{todonotes}
\usepackage{subcaption}
\usepackage{booktabs, makecell, tabularx}
\usepackage{hhline}
\usepackage{comment}

\title{Eclectic Rule Extraction for Explainability\\ of Deep Neural Network based\\ Intrusion Detection Systems}

\author{\IEEEauthorblockN{Jesse Ables\IEEEauthorrefmark{1}, Nathaniel Childers\IEEEauthorrefmark{1}, William Anderson\IEEEauthorrefmark{1}, \\ Sudip Mittal\IEEEauthorrefmark{1}, Shahram Rahimi\IEEEauthorrefmark{1}, Ioana Banicescu\IEEEauthorrefmark{1}, and Maria Seale\IEEEauthorrefmark{2}}

\IEEEauthorrefmark{1} Department of Computer Science \& Engineering \\ Mississippi State University, Mississippi, USA, \\(email: \{jha92, nac294, wha41\}@msstate.edu, \{mittal, rahimi, ioana\}@cse.msstate.edu)\\
\IEEEauthorrefmark{2} U.S Army Engineer Research and Development Center \\ Vicksburg, Mississippi, USA, (email: maria.a.seale@erdc.dren.mil)

}

\begin{document}

\maketitle

\begin{abstract}
    
This paper addresses trust issues created from the ubiquity of black box algorithms and surrogate explainers in Explainable Intrusion Detection Systems (X-IDS). While Explainable Artificial Intelligence (XAI) aims to enhance transparency, black box surrogate explainers, such as Local Interpretable Model-Agnostic Explanation (LIME) and SHapley Additive exPlanation (SHAP), {are difficult to trust}. The black box nature of these surrogate explainers makes the process behind explanation generation {opaque and} difficult to understand. To avoid this problem, one can use transparent white box algorithms such as Rule Extraction (RE). There are three types of RE algorithms: pedagogical, decompositional, and eclectic. Pedagogical methods offer fast but untrustworthy white-box explanations, while decompositional RE provides trustworthy explanations with poor scalability. This work explores eclectic rule extraction, which strikes a balance between scalability and trustworthiness. By combining techniques from pedagogical and decompositional approaches, eclectic rule extraction leverages the advantages of both, while mitigating some of their drawbacks. The proposed Hybrid X-IDS architecture features eclectic RE as a white box surrogate explainer for black box Deep Neural Networks (DNN). The presented eclectic RE algorithm extracts human-readable rules from hidden layers, facilitating explainable and trustworthy rulesets. Evaluations on UNSW-NB15 and CIC-IDS-2017 datasets demonstrate the algorithm's ability to generate rulesets with 99.9\% accuracy, mimicking DNN outputs. The contributions of this work include the hybrid X-IDS architecture, the eclectic rule extraction algorithm applicable to intrusion detection datasets, and a thorough analysis of performance and explainability, demonstrating the trade-offs involved in rule extraction speed and accuracy. 

\end{abstract}

\section{Introduction}
The ubiquity of black box algorithms and black box surrogate explainers create trust issues for Explainable Intrusion Detection Systems (X-IDS) \cite{chennam2022black}. Explainable Artificial Intelligence (XAI) was created as a means to increase the transparency of these black box approaches \cite{szczepanski2020achieving}. Historically, white box techniques were used to create explanations for black box models. More recently, the use of surrogate explainers, such as Local Interpretable Model-Agnostic Explanations (LIME) \cite{Ribeiro2016} or SHapley Additive exPlanations (SHAP) \cite{Lundberg2017}, have become more common. These techniques are used to create local and global explanations for neural networks but are themselves black boxes. By using these techniques, we take a step back in explainability. If one cannot trust a black box model because it is opaque, how can one trust a black box surrogate explainer?

One solution to this problem is the use of pedagogical, white box explanations for neural networks. Pedagogical algorithms take a similar approach to black box surrogate explainers. They use the neural network inputs and outputs to create an approximate model \cite{andrews1995survey, augasta2012rule}. A popular technique is to train Decision Trees (DT) as a surrogate model. Pedagogical approaches have the benefit of being fast and scalable. This method, however, is also lacking with regard to trust. Since pedagogical methods do not use the black box neural network weights, they cannot create a trustworthy surrogate model. Decompositional Rule Extraction (RE) can alleviate this issue. By training DTs using the weights from each layer \cite{andrews1995survey, augasta2012rule}, we can create trustworthy rules for black box neural networks. While decompositional RE gains in trustability, it loses in scalability. A major issue with this type of algorithm is its exponential scaling due to its need to stitch each layer's rules together from input to output. 

Another option is to use eclectic rule extraction. Eclectic RE uses techniques from both pedagogical and decompositional algorithms \cite{augasta2012rule}. Eclectic algorithms offer a middle ground between the scalability of pedagogical techniques and the trustworthiness of decompositional techniques. It trains one or more DTs for each hidden layer which are used to extract rules for a ruleset. Due to this, eclectic rule extraction scales polynomially with respect to the number of layers. This scaling issue can be mitigated using the eclectic algorithm's customizability. For larger Deep Neural Networks (DNN), one can extract rules from a subset of layers rather than all layers. Additionally, the eclectic rule extractor gains the benefit of trust from generating rulesets using weights from the black box neural network. This makes it more trustworthy than black box surrogates and pedagogical approaches.

X-IDS heavily benefits from explainability and trust \cite{neupane2022explainable}. Explainability allows security experts to understand how and why their IDS is making predictions. Experts can use eclectic RE as a means to understand their IDS by generating global, explainable rulesets. Using this information, security experts are able to make modifications to their IDS in order to increase its accuracy. Additionally, experts have other tasks that they need to perform to protect their systems. Having more trust in the IDS can help experts perform tasks in a more timely and confident manner. This leads to more reliable network defense.

In this work, we present a hybrid X-IDS architecture that uses white box eclectic RE to generate explanations. The proposed solution is a white box surrogate explainer that utilizes the DNN's hidden layers to generate an explainable ruleset. DNN models are trained using the UNSW-NB15 and CIC-IDS-2017 datasets, and explainable rulesets are created using the eclectic RE algorithm. We find that the RE algorithm is able to generate rulesets that mimic the models' outputs at an accuracy of 99.9\%. Additionally, the rulesets have similar accuracy to the DNN models when compared to the datasets' ground truth labels.

Major contributions presented in this work are -

\begin{itemize}
    \item A hybrid X-IDS architecture using a black box DNN predictor and a white box surrogate explainer. Eclectic RE is used to generate human-readable rules from the hidden neurons of a DNN. RE creates a global, explainable ruleset that can be used to help users understand how and why their model makes predictions.

    \item An eclectic rule extraction algorithm that can be run on intrusion detection datasets. This algorithm can be run for both binary and categorical predictions. An eclectic RE algorithm gives the user flexibility when determining how much of the model they would like to explain. This can increase ruleset extraction speed. Rulesets generated using this algorithm are highly similar to the DNNs' predictions.
    
    \item A performative and explanatory analysis of our architecture using modern datasets. Our model is tested against the CIC-IDS-2017 and UNSW-NB15 datasets. Using these datasets, we train and test our DNN and create accurate rulesets. Rulesets are able to mimic the DNNs' outputs with an accuracy of 99.9\%. We discuss the trade-off of speed and performance and detail the rulesets' explainability. 
\end{itemize}

This paper is outlined as follows - Section \ref{sec: Related Work} details related works and background information in IDS, XAI, and X-IDS. In Section \ref{sec: rule extraction}, we describe rule extraction techniques and the algorithm used for our experiments. Section \ref{sec: Architecture} explains our hybrid X-IDS architecture. Next, Section \ref{sec: Experimental Results} we discuss our experiment and experimental results. Finally, we detail future works and conclude the paper in Section \ref{sec: conclusion}.

\section{Related Work and Background}
\label{sec: Related Work}

 Here we present various related works and necessary background information pertaining to XAI and X-IDS. This section aims to summarize the current state of explainability in IDS-related materials and contextualize the work described in the following sections.

\subsection{Intrusion Detection Systems (IDS)}
The authors of \cite{ables2022creating} and \cite{denning1987intrusion} define an intrusion as actions that obtain unauthorized access to a network or system. These intrusions can be categorized by their ability to affect a system's Confidentiality, Integrity, and Availability (CIA) \cite{neupane2022explainable}. IDS are tasked with defending a network or host system from intrusion, thereby protecting the CIA principles of security.
In the past, intrusion detection methods took a signature-based approach. The effectiveness of signature-based approaches is determined by the software’s ability to establish, and store, valid malware signatures and check inputs against those signatures \cite{khraisat2019survey}. This method, however, has difficulty detecting \textit{zero-day} attacks. The alternative to signature-based methods is anomaly-based detection. Rather than relying on a database of known signatures, anomaly-based methods use ML algorithms to learn observed behaviors in data. Any behaviors that deviate from the learned behavior is considered an anomaly \cite{khraisat2019survey}.

Many modern IDS systems and research use anomaly-based detections in order to achieve higher accuracy. These approaches can be divided into black box and white box \cite{neupane2022explainable}. Black box algorithms are generally considered difficult to explain and have an opaque decision process \cite{loyola2019black, guidotti2018survey}. These properties make it difficult to discern how and why the model links certain inputs with predicted outputs. Examples of black box IDS used in the literature include Extreme Gradient Boost (XGBoost) \cite{bhati2021improved}, Support Vector Machines (SVM) \cite{kotpalliwar2015classification}, auto-encoder+LSTM neural network \cite{mushtaq2022two}, and 1-Dimensional CNN \cite{qazi2022one}. White box algorithms, on the other hand, are easy to understand and transparent in their decision process. However, they are often less accurate than black box methods due to the simplicity in their design. Examples of white box IDS are Self Organizing Maps (SOM) \cite{ables2022creating} and decision trees \cite{ahakonye2023scada}.

\subsection{Explainable Artificial Intelligence (XAI)}
DARPA defines explainable systems as models that can explain their \textbf{reasoning} to a human, describe the model's \textbf{strengths} and \textbf{weaknesses}, and convey a sense of \textbf{future behavior} \cite{gunning2019darpa}. Using these criteria, explainable models can be created that promote model fairness, privacy, reliability, causality, and trustability \cite{ables2022creating}. 

Similar to IDS and other AI applications, XAI methods fall into black box and white box explanation techniques. Many modern XAI-enabled neural networks feature the use of black box explanation modules. Black box explainers such as Local Interpretable Model-agnostic Explanations (LIME) \cite{Ribeiro2016}, SHapley Additive exPlanations (SHAP) \cite{Lundberg2017}, and Layer-wise Relevance Propagation (LRP) \cite{montavon2019layer}. These methods form associations between input data and predicted outputs, typically using feature importance as a means of explanation. Unfortunately, these method suffer from a lack of trustability \cite{ables2023explainable}. White box explainers can be used to avoid this problem. As mentioned previously, their transparent and simple-to-understand nature allows them to create trustworthy explanations. In our previous works \cite{ables2022creating, ables2023explainable}, we utilize self organizing maps, growing self organizing maps, and growing hierarchical self organizing maps in order to create accurate, highly explainable intrusion detection systems. Explanations are innate to these models, unlike black box neural networks. In order to explain black box neural networks with white box techniques, one must delve into hybrid X-IDS.

\subsection{Hybrid XAI and X-IDS}
Hybrid X-IDS are systems that use black box models to make predictions and white box models to generate explanations. One method found in the literature is to use the inputs and outputs of a NN to train a Decision Tree (DT) classifier. Explainer algorithms such as TREEPAN \cite{craven1995extracting} and HYPINV \cite{saad2007neural} can be used to generate DTs from trained NNs. The authors of \cite{szczepanski2020achieving} use a similar approach to these algorithms to create an X-IDS. They train a set of DTs using the inputs to the NN. Then, the outputs of the NN are used to generate explanations by finding the best DT. Notably, using these methods to explain a NN comes with a significant downside. One is maintaining the black box nature of the NN. Additionally, the proposed methods are designed using white box algorithms, while they are used as if they are black boxes.

Some approaches use the NN's weights in order to generate DTs or rulesets. DeepRED \cite{zilke2016deepred} and ECLAIRE \cite{zarlenga2021efficient} are both extraction algorithms that ``open" the black box to generate explanations. Both of these algorithms are designed to work on DNNs and can generate accurate DTs. Rulesets can then be extracted from the DTs that can be read by humans. Since these algorithms utilize the NN's inputs, weights, and outputs, one may be able to conclude that these algorithms generate more trustworthy explanations. These explainers also work in a white box manner making the explainers themselves trustworthy. The authors of \cite{almutlaq2022two} use DeepRED to generate explanations for various IDS datasets. They note that DeepRED requires a large amount of data to generate accurate rulesets. The creators of ECLAIRE \cite{zarlenga2021efficient} test their against many smaller datasets. They show that their algorithm can generate explainable rulesets. However, their implementation is not well suited for use on large IDS datasets. Additionally, these works compare the rulesets accuracy to the dataset's labels. DTs and rulesets are generated to explain the NN. Our work adds a metric that compares the model's outputs to the dataset's outputs which demonstrates that our extraction algorithm is able to mimic the DNN's thought process.

\section{Rule Extraction in Intrusion Detection}
\label{sec: rule extraction}

In this section, we detail different rule extraction techniques. Rule extraction, being a white box surrogate technique, has some useful benefits that make it a good alternative to black box surrogate explainers. 

\subsection{Rule Extraction Techniques}

Rule Extraction (RE) algorithms are a family of techniques used to generate textual rulesets from neural networks. RE can be categorized into three families: decompositional, pedagogical, and eclectic \cite{hailesilassie2016rule}. The decompositional approach opens up the black box and uses neuron weights and activations to generate rules. This approach generates rules layer-to-layer starting from the output by connecting it to the final hidden layer. It then proceeds in reverse by connecting each hidden layer to the next one, finally connecting the first hidden layer to the model's input layer. A substitution algorithm is used to create a ruleset from this chain of rules. Conversely, the pedagogical approach maintains the model's black box nature. It maps model inputs to outputs and uses this mapping to train a DT. Notably, this approach typically trains only a single tree. Lastly, there is the eclectic approach. Eclectic algorithms use a mixture of decompositional and pedagogical techniques. In this type of algorithm, each hidden layer is used to generate its own textual ruleset. These rulesets can be concatenated together to explain the whole of the network. In this work, we chose to use an eclectic-type RE algorithm. The customizable nature of the algorithm allows us to make design choices that benefit IDS datasets.


Eclectic algorithms lead themselves to be more useful for intrusion detection and its large datasets. Decompositional algorithms such as DeepRed \cite{zilke2016deepred} have an exponential runtime complexity \cite{zarlenga2021efficient}. This complexity is due to how this algorithm substitutes rules from the input layer to the output layer. These algorithms lack flexibility. One must create a ruleset by linking the output to the input, greatly increasing the number of possible substitutions needed to create a ruleset. This in combination with large IDS datasets can cause ruleset generation time to become {unfeasible} for intrusion detection. The eclectic algorithm, however, has the ability to mitigate this problem. Rather than generating rules by connecting all layers, the eclectic algorithm can use a subset or selection of layers. Additionally, the size of the DTs that are created with this method can be adjusted to speed up rule generation.

 \begin{algorithm}
 
 \caption{DNN Rule Extraction}
  \label{alg:DNNRE}
 \begin{algorithmic}[1]
 \renewcommand{\algorithmicrequire}{\textbf{Input: }}
 \renewcommand{\algorithmicensure}{\textbf{Output:}}
 \REQUIRE Dataset ($X$), Model ($M$), Decision Tree Algorithm ($DT_{alg}$)
 \ENSURE  Final Ruleset ($R$)
 \\
 \textbf{BEGIN}
 \\
  \STATE $R = set()$
  \STATE $Y' = \text{M.predict(X)}$
  \FOR{hidden layer $h_i$ in M}
        \STATE $X' = h_i(X)$
        \STATE $hidden_{dt} = DT_{alg}(X', Y')$
        \STATE $Rules_{hidden} = \text{ExtractRules}(hidden_{dt})$
        \STATE $\hat{Y} = Rules_{hidden}(X')$
        \STATE $input_{dt} = DT_{alg}(X, \hat{Y})$
        \STATE $R.add(\text{ExtractRules}(input_{dt}))$
        
  \ENDFOR
\\
 \RETURN $R$ 
 \\
 \textbf{END}
 \end{algorithmic} 

 \end{algorithm}

\begin{figure*}[!ht]
    \centering
    \includegraphics[scale=.75]{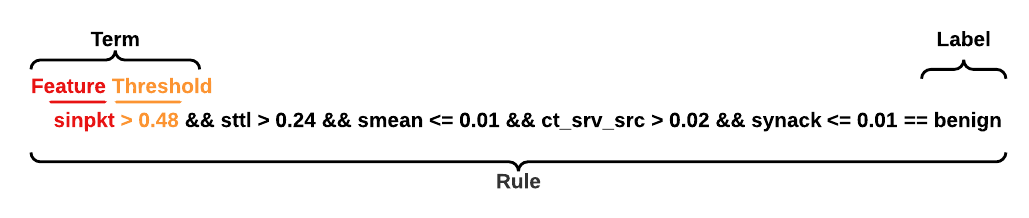}
    \vspace*{-3mm}
    \caption{The composition of a rule. A rule consists of terms that are concatenated together to form a rule. Rules can then be combined in a set to form a ruleset.}
    \label{fig: rule}
\end{figure*}

The pseudo-code for the algorithm we implemented can be found in Algorithm \ref{alg:DNNRE}. The algorithm takes input of a dataset ($X$), a DNN model ($M$), and a decision tree algorithm ($DT_{alg}$). It trains two decision trees per layer that we can extract rules from to generate a ruleset. The algorithm begins by initializing an empty set ($R$) that will be used to store future rules. We then used the trained model to generate predicted labels for our dataset ($Y'$). We then loop through each hidden layer ($h_i$). For each hidden layer, we generate a new dataset ($X'$). This dataset is generated by obtaining the hidden values created by the hidden neurons on each layer. We can then use our hidden value dataset ($X'$) and our predicted labels ($Y'$) to train our hidden value decision tree ($hidden_{dt}$). After the decision tree is trained, we can extract the rules from the tree using a depth-first search approach (ExtractRules()). Once we have our rules extracted, we use the rules to generate a list of labels ($\hat{Y}$). These labels are used in conjunction with the original dataset to create the final \textit{input-to-output} decision tree ($input_{dt}$). The same depth-first search algorithm is used to extract the rules which are then added to the ruleset ($R$). More information about our specific implementation can be found in Section \ref{sec: Explanation Generation}.

\subsection{Rules and Rulesets} 

As discussed earlier, RE algorithms basically take a model and a dataset as an input, and produce a ruleset. Rulesets consist of rules that are used to categorize training or testing samples. Figure \ref{fig: rule} demonstrates the composition of rules and rulesets. Rules are a conjunction of terms. A rule returns true if all of the terms return a true value. When a rule returns true, a categorical label is given to the sample. As the name implies, a ruleset is a set of rules. All rules in a ruleset have unique term compositions. Similarly, all terms in a rule are unique. Terms, in short, are nodes (decisions) on a DT. Rules are paths through the decision tree that are labeled with a leaf node.

\section{Surrogate Explanation X-IDS Architecture}
\label{sec: Architecture}

\begin{figure*}[htbp]
    \centering
    \includegraphics[scale=.5]{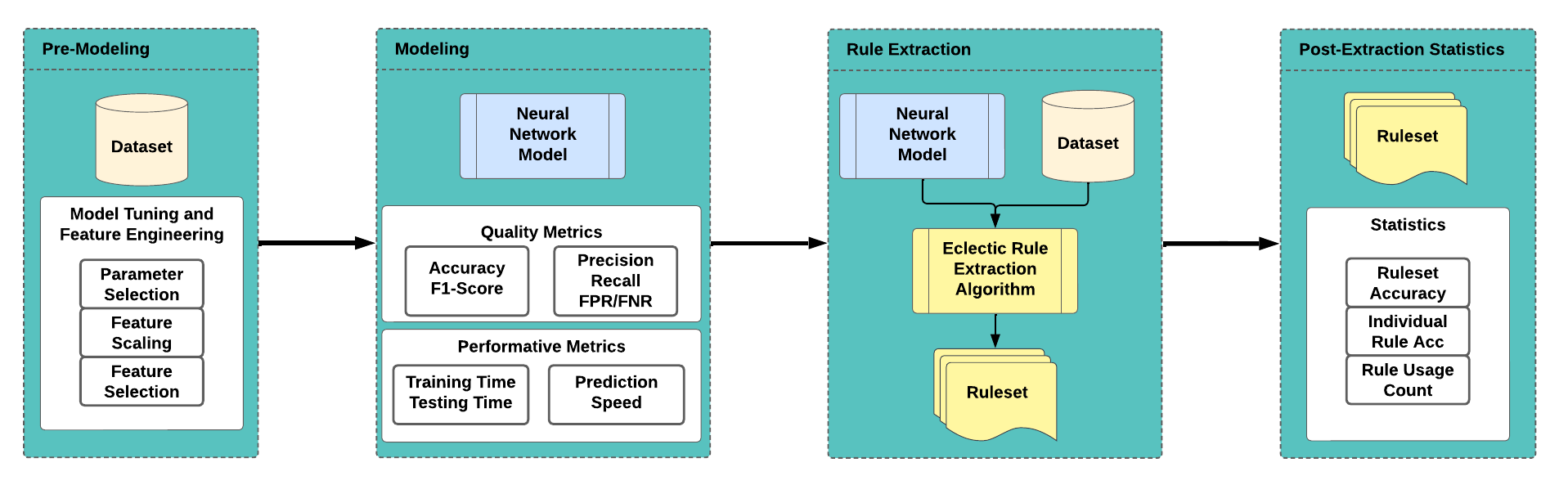}
    \vspace*{-3mm}
    \caption{Architecture for a surrogate explainer X-IDS. It features four total phases. In the Pre-Modeling phase, the datasets are feature engineered to be compatible with the neural network, and model parameters can be selected. The model is trained and tested in the Modeling phase. Here we record important quality and performative metrics. The trained model and dataset can then be used to extract a ruleset. Lastly, we generate statistics for the ruleset and rules to aid the user in their understanding.}
    \label{fig: architecture}
\end{figure*}

One important goal of X-IDS is to aid users in understanding predictions that can help them with certain tasks such as recognition of possible false-positives. To help CSoC (Cyber Security Operation Center) security analysts in their tasks and to protect their networks, we propose a hybrid X-IDS architecture that uses a DNN to create predictions and eclectic RE techniques to create explanations. The architecture is divided into four phases. First, the datasets are preprocessed and model parameters are tuned in the Pre-Modeling phase. Second, the DNN are trained and various quality metrics are recorded. Third, rules are extracted from the model to form rulesets. Fourth, the rulesets are tested for various statistical measures. The architecture diagram for our X-IDS can be found in Figure \ref{fig: architecture}.

\subsection{Pre Modeling}
The first phase in the architecture is Pre Modeling. Here, we construct high-quality datasets, from the input data, and determine model hyper-parameters. This work uses the CIC-IDS-2017 \cite{Panigrahi2018} and UNSW-NB15 \cite{moustafa2015unsw} datasets. There are a number of reasons why we chose these two datasets. First, these datasets use more modern attacks when compared to older datasets. CIC-IDS-2017 and UNSW-NB15 were developed in 2017 and 2015, respectively. Both of these datasets were created to offer `up-to-date' attacks. Although these datasets are six years old, they can be used to give a good impression of how our model will work with real-world data. Second, these datasets contain millions of samples. 
{The larger datasets can show how the algorithm can be used with real world data and can demonstrate the algorithms' scalability.} Understanding the scalability of our model, while explainable, is a crucial factor for intrusion detection. In addition, we are able to create large validation and testing datasets that contain data which may not appear in the training dataset.

We use a few techniques to preprocess the datasets. First, we begin by removing samples with empty values. Then we can one hot encode the categorical features. Finally, we normalize the values. During this phase, we can also choose to feature engineer the datasets. Lastly, we tune model parameters to achieve higher accuracies. This includes many common NN options. The number of hidden layer neurons and the number of training iterations are the primary options we change to create an accurate IDS model. The general make-up of the NN can be found in the following section.

\subsection{Modeling}

The next phase is Modeling. In this phase, we train the black box neural network and record various quality and performative metrics. To construct the NN, we use Tensorflow \cite{tensorflow2015-whitepaper}. It consists of an input layer, two r hidden layers, and an output layer. The two hidden layers each contain 64 neurons and use the \textit{ReLU} activation function. The output layer uses the \textit{Sigmoid} activation function for binary classification. To optimize the model, we selected the \textit{Adam} optimizer. After the model has been trained, it can be tested for various quality and performative metrics. Table \ref{table:nn parameters} shows the DNN parameters we used during training.

\begin{table}[!htb]
    \small
    \setlength{\tabcolsep}{8pt}
\begin{tabularx}{\linewidth}{
       >{\centering\arraybackslash}p{0.15\textwidth}|X
       >{\centering\arraybackslash}X}
        \hhline{|=|=|}
\thead{\textbf{Parameter Name}}  & \thead{\textbf{Parameter Value}} \\
        \hline
    \makecell{Hidden Layers} &
    \makecell{2} \\
    
    \hline
    \makecell{Hidden Layer\\Neuron Count} &
    \makecell{64} \\
    
    \hline
    \makecell{Output Layer\\Neuron Count} &
    \makecell{2} \\
    
    \hline
    
    \makecell{Hidden Layer\\ Activation Function} &
    \makecell{Rectified Linear\\Unit (ReLU)} \\
    
    \hline
    \makecell{Output Layer\\ Activation Function} &
    \makecell{Softmax} \\
    
    \hline
    \makecell{Bactch Size} &
    \makecell{64} \\
    
    \hline
    \makecell{Training Iterations} &
    \makecell{100} \\
    
    \hline
    \makecell{Early Stopping\\(Validation Loss)} &
    \makecell{5} \\
    
    \hline

        \hhline{|=|=|}
    \end{tabularx}
    \caption{The parameters used for training our neural network model. All layers are fully connected and use the ReLU activation function. The eclectic rule extraction algorithm we implemented requires there to be at least two output neurons. Two output neurons can effectively be used for binary classification with the softmax activation function.}
    \label{table:nn parameters}
\end{table}

\textit{Quality Metrics:} For our experiments, we record many traditional quality metrics. These include accuracy, F1-score, precision, recall, False Positive Rate (FPR), and False Negative Rate (FNR). Accuracy compares the number of true positives and negatives to the number of false positives and negatives. This gives a general idea of how well the model performs as a whole. However, its use may be misleading due to imbalanced datasets. The F1-score, on the other hand, accounts for this by using precision and recall to define its score. This helps to minimize the effects of an imbalanced dataset. The last two metrics are FPR and FNR. These detail how often the model mislabels anomalous data as normal and vice-versa. These are important metrics for an IDS as it denotes how often an attack goes unnoticed or a benign user is prevented from using a service.

\textit{Performative Metrics:} There are also performance-based metrics that are important to note for an IDS. These include training, testing, and prediction times. The speed at which an IDS can be trained and tested can be vital for a network.

\subsection{Rule Extraction}
\label{sec: Explanation Generation}

In this stage, we use the trained model and the training dataset to extract rules from the model's hidden layers. We outline the eclectic rule extraction algorithm in Section \ref{sec: rule extraction}. Rules generated from the hidden layer can be concatenated together to form an explainable ruleset. This ruleset can be used by the user to understand the potential decisions the model is making when determining if a sample is benign or malicious.

The rule extraction algorithm depends on the use of a Decision Tree (DT). We chose to use the Scikit-learn Decision Tree Classifier. However, there are other DT classifiers that are available, such as C5.0, that offer varying functionality and scalability. Scikit-learn's DT offers the benefit of speed and ease of use, which is the reason why we chose their implementation. In addition, the DTs come with varying hyper-parameters that can be used to alter the training process. Notably, the `max depth' and `max leaves' hyper-parameters can be used to limit the size and number of rules generated. We can modify these parameters to find the optimal trade-off between speed and ruleset accuracy. Lastly, it is possible to modify the DT training process by changing the amount of data they train on. The main scalability factor with this eclectic RE algorithm is dataset size. One could decide to only use a subset of the original training dataset in order to speed up the rule extraction process. This comes with its own downsides, however. By limiting the amount of data the DTs are trained on, one may be leaving out vital information for an accurate ruleset.

There are some additional notes that can be made about our specific implementation. First, the algorithm is designed to work with multiclass datasets. This means that NNs need to have at least two output neurons. Multiclass classification can abstractly predict binary classes by using two output neurons and the softmax activation function. The user will need to One Hot Encode their binary class dataset. Secondly, we did not implement multiprocessing. As mentioned later in Section \ref{sec: Experimental Results}, the bottleneck for our implementation is Python's default single thread. Implementing the ability to use more than one CPU core should drastically increase the speed of the algorithm.

\begin{table*}[ht]
\resizebox{\textwidth}{!}{%
\begin{tabular}{lcccccccc}
\multicolumn{9}{c}{\textbf{UNSW-NB15}}                                                                  \\ \hline
\textbf{Experiment} &
  \textbf{\begin{tabular}[c]{@{}c@{}}Num.\\Rules \end{tabular}} &
  \textbf{\begin{tabular}[c]{@{}c@{}}Ground Truth\\Accuracy \end{tabular}} &
  \textbf{\begin{tabular}[c]{@{}c@{}}Model Prediction\\ Accuracy \end{tabular}} &
  \textbf{\begin{tabular}[c]{@{}c@{}}Average\\Terms \end{tabular}} &
  \textbf{\begin{tabular}[c]{@{}c@{}}Longest\\Rule \end{tabular}} &
  \textbf{\begin{tabular}[c]{@{}c@{}}Extraction\\Time (s) \end{tabular}} &
  \textbf{\begin{tabular}[c]{@{}c@{}}Testing\\Time (s) \end{tabular}} &
  \textbf{\begin{tabular}[c]{@{}c@{}}Testing\\Std (s) \end{tabular}} \\ \hline
\textbf{Unbounded}  & 2380 & 93.6\% & 99.1\% & 15.8 & 30 & 1610 & 723 & 79 \\
\hline
\textbf{2000 Leaves}& 2421 & 93.6\% & 99.0\% & 15.7 & 30 & 1600 & 805 & 140\\
\textbf{1000 Leaves}& 1684 & 93.7\% & 99.1\% & 14.7 & 25 & 1603 & 549 & 95 \\
\textbf{500 Leaves} & 946  & 93.6\% & 99.0\% & 13.1 & 25 & 993  & 278 & 65 \\
\textbf{100 Leaves} & 184  & 93.5\% & 98.5\% & 4.2  & 14 & 212  & 51  & 7  \\
\textbf{10 Leaves}  & 19   & 90.9\% & 94.7\% & 4.2  & 5  & 60   & 6.6 & .5 \\
\hline
\textbf{20 Layers}  & 2278 & 93.6\% & 99.0\% & 20   & 20 & 1532 & 606 & 98 \\
\textbf{10 Layers}  & 502  & 93.4\% & 98.4\% & 10   & 10 & 430  & 89  & 10 \\
\textbf{5 Layers}   & 38   & 91.0\% & 95.1\% & 5    & 5  & 77   & 11  & 2  \\
\hline\hline
\multicolumn{9}{c}{\textbf{CIC-IDS-2017}}                                                             \\ \hline
 \textbf{Experiment} &
  \textbf{\begin{tabular}[c]{@{}c@{}}Num.\\Rules \end{tabular}} &
  \textbf{\begin{tabular}[c]{@{}c@{}}Ground Truth\\Accuracy \end{tabular}} &
  \textbf{\begin{tabular}[c]{@{}c@{}}Model Prediction\\ Accuracy \end{tabular}} &
  \textbf{\begin{tabular}[c]{@{}c@{}}Average\\Terms \end{tabular}} &
  \textbf{\begin{tabular}[c]{@{}c@{}}Longest\\Rule \end{tabular}} &
  \textbf{\begin{tabular}[c]{@{}c@{}}Extraction\\Time (s) \end{tabular}} &
  \textbf{\begin{tabular}[c]{@{}c@{}}Testing\\Time (s) \end{tabular}} &
  \textbf{\begin{tabular}[c]{@{}c@{}}Testing\\Std (s) \end{tabular}} \\ \hline
\textbf{Unbounded}  & 1686& 93.1\% & 99.9\% & 14.2 & 27 & 9504 & 6882 & 1321 \\
\hline
\textbf{2000 Leaves}& 1815 & 93.1\% & 99.9\% & 14.1 & 27 & 9782 & 6211& 566\\
\textbf{1000 Leaves}& 1701 & 93.1\% & 99.9\% & 14.3 & 27 & 8969 & 4950& 879\\
\textbf{500 Leaves} & 1000 & 93.1\% & 99.9\% & 12.2 & 25 & 7276 & 3305& 719\\
\textbf{100 Leaves} & 188  & 93.0\% & 99.8\% & 8.6  & 18 & 1792 & 554 & 129\\
\textbf{10 Leaves}  & 20   & 91.5\% & 97.6\% & 4.7  & 6  & 664  & 67  & 11 \\
\hline
\textbf{20 Layers}  & 1675 & 93.1\% & 99.9\% & 13.2 & 20 & 4401 & 4659& 1409\\
\textbf{10 Layers}  & 601  & 93.0\% & 99.6\% & 9.1  & 10 & 2060 & 1242& 301 \\
\textbf{5 Layers}   & 57   & 91.0\% & 97.0\% & 4.9  & 5  & 760  & 113 & 30  \\
\hline\hline
\end{tabular}%
}
\caption{This table shows the results from the unbounded, leaves, and layers tests. Tests were run on DNN that achieved an accuracy of 93.7\% for UNSW-NB15 and 93.1\% for CIC-IDS-2017. The unbounded experiment allows the rule extraction algorithm to generate trees with unlimited leaves and layers. The leaves and layers experiments limit the number of leaves and layers the decision tree algorithm is allowed to produce. This speeds the rule extraction up at the cost of accuracy. Accuracy is measured in two ways: ground truth accuracy and model prediction accuracy. Ground truth accuracy compares the ruleset's prediction ability to the testing dataset's true labels. Model prediction accuracy compares the ruleset's prediction accuracy to the model's predictions.}
\label{table: leaves and layers}
\end{table*}

\subsection{Post-Extraction Statistics}
Finally, we can compare the dataset and model predictions to the ruleset to obtain useful statistical data. Useful information includes ruleset accuracy, individual rule accuracy, and rule usage. Using these statistics, we can aid the user in understanding the ruleset and model. The first step in this process is ruleset evaluation.

There are some major design decisions that can be made when evaluating a ruleset. First, one can opt to take a comprehensive or greedy approach to evaluating the ruleset. We opted to use a greedy rule comparison approach. The difference between these approaches is their stopping criteria for each testing sample. In the comprehensive approach, each sample is compared to all rules. Since there is a potential for collision, an additional step would need to be made to determine which rule is more accurate. The second method is a greedy approach. Rather than compare a sample to all possible rules, we stop as soon as we find a matching rule. The benefit of this approach is that it speeds up the evaluation on average by a factor of 2. This is due to the fact that on average a sample should only need to be tested against half the rules. This, however, does not change the potential maximum runtime.

A potential problem with RE is the number of rules it extracts. Some extraction algorithms can extract 10,000 rules. Our algorithm has extracted close to 2400 rules. These large number of rules are an unreasonable amount for a human to process. To mitigate this, we can assign a usage counter and accuracy to each rule. During ruleset evaluation, these statistics can be saved in order to assist the user in understanding the ruleset. Higher used and higher accuracy rules can be used to understand the general composition of benign or malicious samples.

\section{Experimental Results}
\label{sec: Experimental Results}

\begin{table*}[ht]
\resizebox{\textwidth}{!}{%
\begin{tabular}{lcccccccc}
\multicolumn{9}{c}{\textbf{UNSW-NB15}}                                                                  \\ \hline
\textbf{Experiment} &
  \textbf{\begin{tabular}[c]{@{}c@{}}Num.\\Rules \end{tabular}} &
  \textbf{\begin{tabular}[c]{@{}c@{}}Ground Truth\\Accuracy \end{tabular}} &
  \textbf{\begin{tabular}[c]{@{}c@{}}Model Prediction\\ Accuracy \end{tabular}} &
  \textbf{\begin{tabular}[c]{@{}c@{}}Average\\Terms \end{tabular}} &
  \textbf{\begin{tabular}[c]{@{}c@{}}Longest\\Rule \end{tabular}} &
  \textbf{\begin{tabular}[c]{@{}c@{}}Extraction\\Time (s) \end{tabular}} &
  \textbf{\begin{tabular}[c]{@{}c@{}}Testing\\Time (s) \end{tabular}} &
  \textbf{\begin{tabular}[c]{@{}c@{}}Testing\\Std (s) \end{tabular}} \\ \hline
\textbf{Unbounded}    & 2380 & 93.6\% & 99.1\% & 15.8 & 30 & 1610 & 723 & 79 \\
\hline
\textbf{80\% Dataset} & 2029 & 93.6\% & 98.9\% & 15.6 & 28 & 1106 & 535 & 44 \\
\textbf{60\% Dataset} & 1735 & 93.6\% & 98.8\% & 14.9 & 28 & 695  & 589 & 89 \\
\textbf{40\% Dataset} & 1377 & 93.4\% & 98.6\% & 14.4 & 27 & 358  & 324 & 23 \\
\textbf{20\% Dataset} & 841  & 93.3\% & 98.3\% & 13.2 & 23 & 110  & 281 & 76 \\
\hline
\textbf{First Hidden} & 1391 & 93.6\% & 99.0\% & 15.5 & 29 & 768  & 561 & 93 \\
\textbf{Second Hidden}& 1396 & 93.6\% & 99.0\% & 15.5 & 30 & 828  & 471 & 63 \\
\hline\hline
\multicolumn{9}{c}{\textbf{CIC-IDS-2017}}                                                             \\ \hline
 \textbf{Experiment} &
  \textbf{\begin{tabular}[c]{@{}c@{}}Num.\\Rules \end{tabular}} &
  \textbf{\begin{tabular}[c]{@{}c@{}}Ground Truth\\Accuracy \end{tabular}} &
  \textbf{\begin{tabular}[c]{@{}c@{}}Model Prediction\\ Accuracy \end{tabular}} &
  \textbf{\begin{tabular}[c]{@{}c@{}}Average\\Terms \end{tabular}} &
  \textbf{\begin{tabular}[c]{@{}c@{}}Longest\\Rule \end{tabular}} &
  \textbf{\begin{tabular}[c]{@{}c@{}}Extraction\\Time (s) \end{tabular}} &
  \textbf{\begin{tabular}[c]{@{}c@{}}Testing\\Time (s) \end{tabular}} &
  \textbf{\begin{tabular}[c]{@{}c@{}}Testing\\Std (s) \end{tabular}} \\ \hline
\textbf{Unbounded}    & 1686 & 93.1\% & 99.9\% & 14.2 & 27 & 9504 & 6882& 1321\\
\hline
\textbf{80\% Dataset} & 1474 & 93.1\% & 99.9\% & 14.2 & 25 & 6964 & 4388& 1599\\
\textbf{60\% Dataset} & 1389 & 93.1\% & 99.9\% & 13.5 & 25 & 4244 & 4630& 1573\\
\textbf{40\% Dataset} & 990  & 93.1\% & 99.9\% & 13.1 & 24 & 2416 & 3208& 1348\\
\textbf{20\% Dataset} & 826  & 93.1\% & 99.9\% & 12.1 & 23 & 877  & 2295& 753 \\
\hline
\textbf{First Hidden} & 905  & 93.1\% & 99.9\% & 14.1 & 27 & 4625 & 4184& 1138\\
\textbf{Second Hidden}& 909  & 93.1\% & 99.9\% & 14.1 & 27 & 4837 & 3509& 636 \\
\hline\hline
\end{tabular}%
}
\caption{This table shows the results from the unbounded, training data subsets, and hidden layer tests. Tests were run on DNN that achieved an accuracy of 93.7\% for UNSW-NB15 and 93.1\% for CIC-IDS-2017. The unbounded experiment allows the rule extraction algorithm to generate trees with unlimited leaves and layers. The data subset tests limit the amount of training data used to extract rules. This speeds up the training time linearly but does not necessarily affect testing time. The hidden layer limitation extracts rules from each hidden layer. Extraction time is cut in half when compared to the unbounded test. Testing time is also increased due to fewer rules extracted. Accuracy is measured in two ways: ground truth accuracy and model prediction accuracy. Ground truth accuracy compares the ruleset's prediction ability to the testing dataset's true labels. Model prediction accuracy compares the ruleset's prediction accuracy to the model's predictions.}
\label{table: subset and hidden}
\end{table*}

We evaluate our X-IDS's eclectic rule extraction technique against a trained DNN. The DNN is trained up to 100 epochs with early stopping. Training ends early when the validation loss does not improve with a patience of 5 epochs. Early stopping is used as a means to speed up training time after the model has reached a minimum. In general, the models would train an average of 25 to 30 epochs before stopping. We trained the DNN on two datasets: CIC-IDS-2017 and UNSW-NB15. These datasets are separated into training (60\%), validation (20\%), and testing (20\%) sets. The datasets are preprocessed for binary classification. However, the label datasets are One Hot Encoded to work with our RE implementation (See Section \ref{sec: Explanation Generation}). Our models achieved an accuracy of 93.1\% on CIC-IDS-2017 and 93.7\% on UNSW-NB15. These accuracies are important to show how effective the eclectic rule extractor is. Extracted rulesets are tested for model prediction and ground truth accuracy. Model prediction accuracy compares the model's predictions to the ruleset's predictions. Ground truth accuracy compares the ruleset's accuracy to the testing dataset's true labels. We define high accuracy for model prediction accuracy as $>$99\% and ground truth accuracy as with 1\% of the model's accuracy.  Rulesets are only extracted once; however, the rulesets are randomized and tested five times. Randomizing the rulesets before testing shows that the greedy rule selection approach is accurate. Because of this, we also record the standard deviation to demonstrate the algorithm's volatility. The experiments aim to demonstrate the trade-off of accuracy to speed and to determine potential optimal parameters.

\subsection{Ruleset Experiments}
Our ruleset experiments are divided into a few different categories. First, the RE algorithm creates rulesets using default, unbounded parameters. This allows all DTs to generate as many layers and leaves as they need. Additionally, we train the RE algorithm using the full training dataset. Second, we create rulesets by limiting the number of layers the DTs can generate. This lowers the total number and size of rules. Third, we limit the number of leaves the DTs can generate. This can allow the DTs to grow as many layers as they want but reduces the number of total rules that can be created. Fourth, we limit the RE algorithm to each layer. Since our models have two hidden layers, we will test how accurate the rulesets are when they are generated from each layer. Last, we will train the RE using subsets of the training dataset. Although each experiment only changes one parameter, optimal parameters could be a mixture of the above changes. 

\begin{figure*}[!ht]
    \hspace{15pt}
     \begin{subfigure}{0.2\textwidth}
         \hspace{15pt}
         \subcaptionbox{UNSW-NB15 layer speed up compared to accuracy loss\label{fig:UNSW layers speedup}}{%
         \includegraphics[scale=.45]{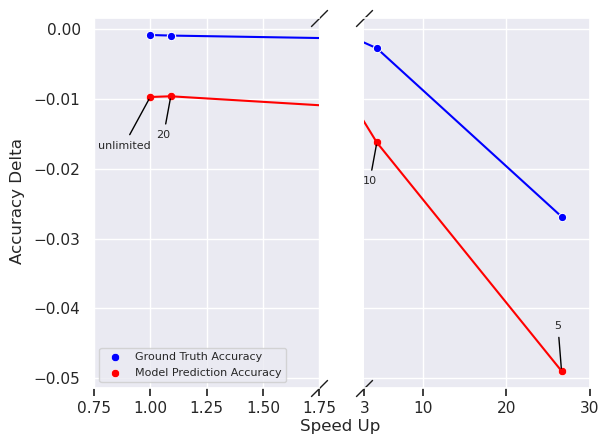}}
         
     \end{subfigure}
     \hspace{130pt}
     \begin{subfigure}{0.3\textwidth}
         \subcaptionbox{UNSW-NB15 leaves speed up compared to accuracy loss\label{fig:UNSW leaves speedup}}{%
         \includegraphics[scale=.45]{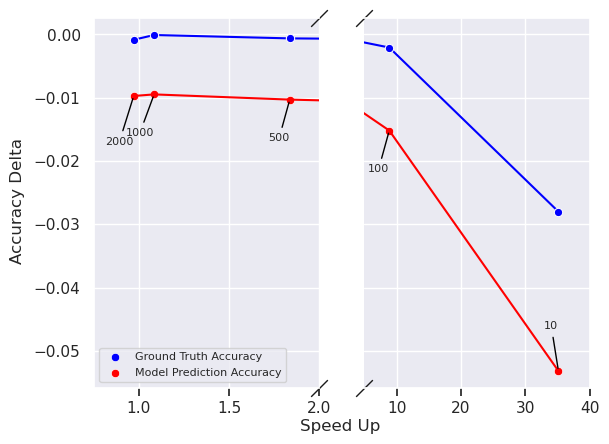}
         }
     \end{subfigure}
     \vfill
     \hspace{10pt}
     \begin{subfigure}{0.3\textwidth}
     \subcaptionbox{\mbox{CIC-IDS-2017 layer speed up compared to accuracy loss}\label{fig:CIC layers speedup}}{%
         \includegraphics[scale=.45]{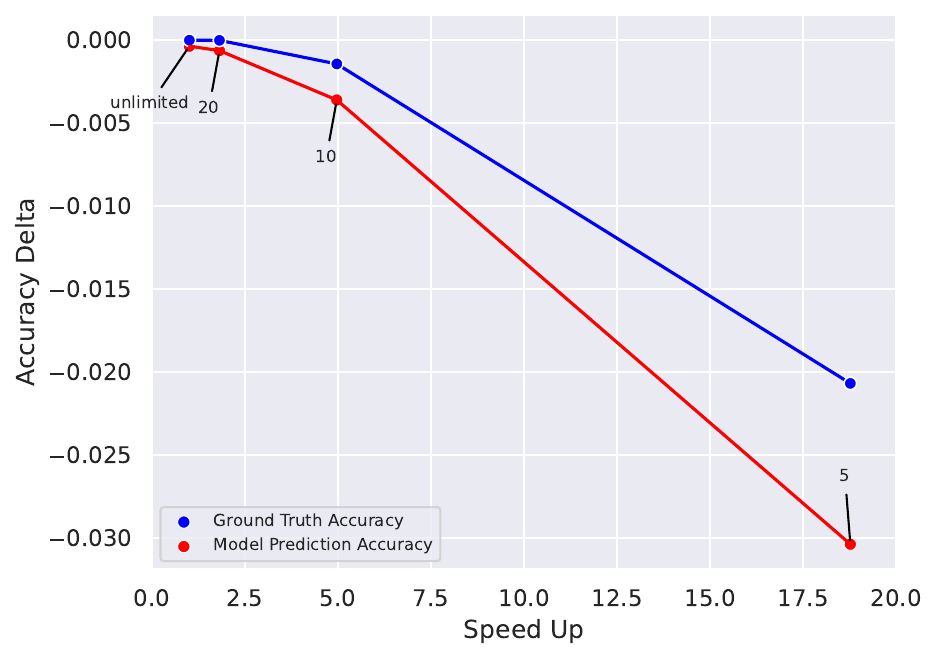}
         }
         
     \end{subfigure}
     \hspace{80pt}
     \begin{subfigure}{0.3\textwidth}
         \subcaptionbox{\mbox{CIC-IDS-2017 leaves speed up compared to accuracy loss}\label{fig:CIC leaves speedup}}{%
         \includegraphics[scale=.45]{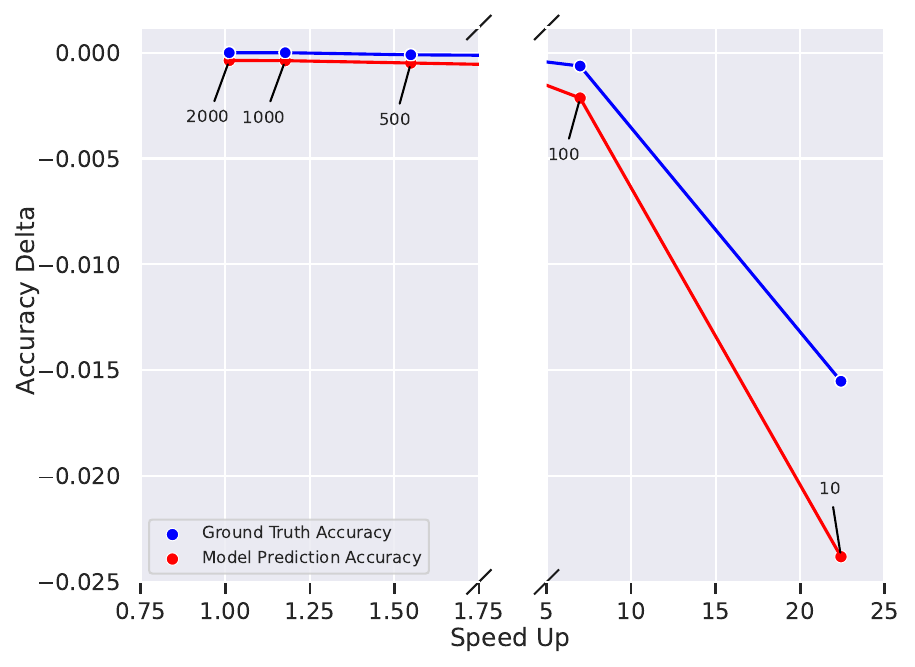}
         }
     \end{subfigure}
        \caption{These charts compare the speed up versus accuracy loss for the UNSW-NB15 and CIC-IDS-2017 rulesets. Ground truth accuracy is the rulesets label versus the testing datasets labels. Model prediction accuracy is the rulesets labels versus the models' predicted outputs. Figures (a)(b)(d) are split into two charts with different scales. Points on each of the figures are labeled with their experiment name.}
\label{fig: speedup}
\end{figure*}

\subsection{Unbounded Eclectic Rule Extraction}
As mentioned, our first experiment used default DT parameters and the full training dataset. This allows the DTs to be trained unbounded. It can generate as many layers and leaves as it needs. Rulesets generated this way were nearly as accurate as their DNN and had high model prediction accuracy. However, a major downside to creating rulesets this way is the amount of time needed and the size of the rules. The UNSW-NB15 model took 1610 seconds to extract and, on average, 723 seconds to test. Testing had a standard deviation of 78.6 seconds. UNSW had a model prediction accuracy of 99.04\% and a ground truth accuracy of 93.6\%. CIC-IDS-2017 was extracted in 9504 seconds (2.6 hours) and tested in 6962 seconds (1.9 hours). This model's testing speed had a standard deviation of 1511 seconds (.4 hours). It had a model prediction accuracy of 99.96\% and a ground truth accuracy of 93.1\%. Both datasets' prediction accuracy and true label accuracy are within one percentage point, which is considered acceptable for this study. As one can see, there is a large deviation in testing times. There are several potential reasons behind this. First, the ruleset can accurately separate data so that few samples have overlapping rules. This can mean many rules must test many samples before finding their match. Additionally, some rules tend to be champions for each label. Depending on where these rules are ordered in the ruleset can drastically change the speed at which the algorithm runs. Lastly, longer rules that are checked before finding the matching rule will also increase runtime.

\subsection{Limited Leaves}
The next set of experiments limits the number of leaves that DTs are allowed to generate. This limits the rulesets in a few ways. First, there is a maximum amount of rules that are allowed to be generated. This number is $\leq n * 2$, where $n$ is the limit of leaves. This phenomena can be seen in \ref{table: leaves and layers}. For both datasets, the 500 leaves and below are only to generate less than or equal to the maximum allowed rules. Some of these experiments do not reach their maximum due to the second hidden layer creating duplicate rules. Second, although there is no explicit limit on the number of layers, limiting the leaves can limit the number of layers. This effect can be seen in the 100 and 10 leaves limited experiments. We see that accuracy is associated with the number of rules; however, there is an upper limit on the number of rules needed for high accuracy. Users are able to limit the RE algorithm greatly before accuracy begins to degrade below our 1\% criteria. UNSW-NB15 can be limited to 500 leaves and still maintain high accuracy. Its speed can be increased further if only 100 leaves are used, but its ability to mimic model output degrades by 0.5\%. CIC-IDS-2017 is able to maintain high accuracy when limited to 100 leaves. Likely, this is due to the large training dataset size. Figures \ref{fig: speedup}, demonstrate the trade-off of accuracy to speed. We define speedup as the combination of unbounded training and average testing time divided by the limited experiment's training and average testing time. Depending on the datset, one can see 5 to 10 times speedup before losing 1\% accuracy.

\subsection{Limited Layers}
Table \ref{table: leaves and layers} also shows the results from the limited layers test. This limit strictly affects the number of layers and implicitly restricts the number of rules. These experiments demonstrate the ability to speed up the algorithms by limiting ruleset creation. The total number of rules for the unbounded and 20 layer experiments are similar. However, the extraction time is reduced by a factor of 2. This is likely due to the decrease of the average length and the longest rule. UNSW-NB15 is able to maintain high accuracy using the 20 layer limitation. However, it loses 0.6\% model prediction accuracy when limited to 10 layers. CIC-IDS-2017 follows a similar trend as in the previous experiment. It is able to have high accuracy even with the more limited parameters. Even with only 5 layers, it is able to mimic the model's predictions with an accuracy of 97\%. Again, its ability to maintain high accuracy when compared to UNSW-NB15 is likely due to the larger amount of training samples.

\subsection{Training Data Subsets}
The training data subset experiments seek to improve speed by limiting the amount of training data used to create rulesets. Generally, we see linear increases in extraction and testing time with respect to dataset size. Here we see that training dataset size is an important factor for model prediction accuracy. Although minor, we see an immediate degradation of model prediction accuracy for UNSW-NB15. Ground truth accuracy is able to maintain high accuracy, but we begin to lose model explainability. On the other hand, CIC-IDS-2017 is able to maintain high accuracy throughout all the subset experiments. Have in mind that 20\% of the CIC-IDS-2017 dataset is still larger than the UNSW-NB15 dataset.

\subsection{Limited DNN Hidden Layers}
The last set of experiments tests how rulesets generated from each layer perform. Using this parameter cuts the number of rules in half and greatly increases the algorithm's speed as the algorithm scales exponentially with respect to the number of hidden layers. However, using this, debatably, limits the ruleset's explainability. Rather than explaining the full model, one is only explaining part of the model. In our case, we are only explaining half of the model. UNSW-NB15 is able to maintain 99\% model prediction accuracy. A user is able to cut the extraction time in half using this limitation while also maintaining high accuracy. We see a similar trend for CIC-IDS-2017. This raises the question: ``Is creating high accuracy rulesets, by extracting from only part of the DNN, less explainable?" Unfortunately, there is no definitive answer to this question. The answer is likely subjective with respect to individual users.

\subsection{Explainability Discussion}
Due to the size of some of the rulesets, it is important to discuss the usability of eclectic rule extraction. Additionally, it is important to discuss the explainability and trustworthiness of certain limited rulesets. Our RE algorithm created as many as 2400 rules when unbounded. Additionally, the unbounded rulesets generated rules with an average of 15.8 terms and a max of 30 terms. These two facts combine to make it a difficult task for humans to parse rulesets. By limiting the algorithm in the various ways above, we are able to decrease the size of the rules and rulesets. This makes the rulesets easier for users to parse but potentially lowers the ruleset's accuracy. With this in mind, we should ask a few questions. First, ``does limiting the DTs decrease the ruleset's explainability and trustworthiness?" Second, ``is model prediction accuracy directly related to explainability and trustworthiness?" Third, ``what methods can users use to understand rulesets?"

The first and second questions are interlinked. The answers to these questions are likely subjective and open to debate. One user may value model prediction accuracy and ground truth accuracy similarity over all other metrics. This is because they are the only concrete statistics that one can use to compare DNN model and ruleset. Limiting ruleset creation would only decrease explainability when accuracy begins to degrade. The question then becomes ``how much can accuracy degrade before this type of user no longer trusts the ruleset?" Another user may value information as a means of determining trustworthiness. Longer rules and rulesets may seem more explainable, especially because these typically correlate with higher accuracy.

The third question can have a more concrete answer. Rulesets are able to record how many and how accurate they are with the testing dataset. Rules can then be sorted by the most used or the most correct. This is applicable for both ground truth and model prediction accuracy. Users can view the most used rules and their labels. These rules can be used to form a general, global understanding of the DNN model. Users may be able to determine which features allow for higher accuracy. Using this information, the user may be able to determine which features should be removed from the dataset to make for more accurate predictions. Additionally, one can use the way DT algorithms train to their advantage. Scikit-learn's DT mainly focuses on information gain. This means that higher-level terms will typically have more variance. These terms will appear in more rules meaning they are more significant than other terms and features. Lastly, it may be possible to use an algorithm to summarize the rulesets. This could be useful on larger rulesets, but it may run into the issue of explaining a white-box with a black-box.

\section{Conclusion and Future Works}
\label{sec: conclusion}


In this paper, we created an X-IDS architecture that uses eclectic rule extraction to generate explanations for a DNN. Our X-IDS created rulesets that were up to 99.9\% accurate when compared to our models' outputs. Our rulesets also had a similar accuracy to the DNN models when compared to the testing datasets true labels. The experiments run demonstrate the eclectic RE algorithm's scalability and customizability. By limiting our rule extraction algorithm, we can greatly increase its speed. However, its accuracy can begin to degrade the more the algorithm is limited. This gives the user the choice between accuracy, explainability, and speed. 

Potential future works include extending the eclectic rule extraction algorithm to recurrent neural networks or other highly accurate AI models. For X-IDS architecture to be trusted, both the model and the explainer need to be accurate. Another potential future work could involve translating extracted rules into directly useful firewall rules. Rather than giving the user a set of rules, the rulesets themselves could be explained by creating firewall rules. 

\section*{Acknowledgement}

\noindent This work by Mississippi State University was financially supported by the U.S. Department of Defense (DoD) High Performance Computing Modernization Program, through the US Army Engineer Research and Development Center (ERDC) (\#W912HZ-21-C0058) and National Science Foundation Award \#2234515. The views and conclusions contained herein are those of the authors and should not be interpreted as necessarily representing the official policies or endorsements, either expressed or implied, of the U.S. Army ERDC or the U.S. DoD.

\bibliographystyle{unsrt}
\bibliography{refs}

\end{document}